\documentstyle[12pt,epsfig]{article}

\begin{document}
  \newcommand{\ccaption}[2]{
    \begin{center}
    \parbox{0.85\textwidth}{
      \caption[#1]{\small{{#2}}}
      }
    \end{center}
    }

\vspace{1truecm}
\begin{center}
{\large { \bf
CHARMED HADRONS PRODUCTION IN HIGH-ENERGY $\Sigma^{-}$ BEAM }}
\end{center}

\vspace*{0.5truecm}

\begin{center}
A.K.~Likhoded$^1$~and~S.R.~Slabospitsky \\
State Research Center \\
Institute for High Energy Physics, \\
Protvino, Moscow Region 142284, \\ RUSSIA
\end{center}

\vspace*{1.5truecm}

\begin{center}
Abstract
\end{center}
We present the calculation of the inclusive $x_F$-distributions of charmed 
hadrons, produced in high-energy $\Sigma^-$-beam. Our calculation is based on 
the modified mechanism of charmed quarks fragmentation as well as on the 
mechanism of $c$-quark recombination with the valence quarks from initial 
hadrons.

\vfill

\rule{3cm}{0.5pt}

$^1$E--mail:~LIKHODED$@$mx.ihep.su,

\newpage

Perturbative QCD provides a reasonable description of the experimental data 
on the inclusive cross sections of open charm and beauty production on fixed
targets~\cite{1}. However, it is well known that the experiments indicate that
there is a substantial difference in the production of the charmed and
anticharmed hadrons in the fragmentation region of the initial hadrons 
(a leading particle effect). This leading particle asymmetry $A$ is defined 
as follows:
\begin{eqnarray}
 A \equiv \frac{ \sigma({\rm leading}) \, - \, \sigma({\rm nonleading})}
  {\sigma({\rm leading}) \, + \, \sigma({\rm nonleading})}.
\label{asym}
\end{eqnarray}
It worth to note, that perturbative QCD calculation~\cite{1} is unable to 
reproduce this effect. 
Indeed, in the quark parton model framework the production of hadrons 
containing a heavy quark proceeds via two subsequent stages 
\begin{itemize}
\item{
heavy quark pair $Q\bar Q$ is produced as a result of the hard
collision of the partons from the initial hadrons (e.g. the subprocesses 
$gg \to c \bar c$ and
$q \bar q \to c \bar c$ in Born approximation); 
} 

\item{
transition of the heavy quarks $c$ into charmed hadrons (``hadronization'').
}
\end{itemize}

The standard way of describing heavy quark  hadronization is to use
the fragmentation function $D(z)$ of the heavy $c$-quark into the charmed
hadron ($D$-meson or baryon)
(here $z = |\vec p_D| / |\vec p_c|$ is the fraction of the heavy quark
momentum carried away by the charmed hadron $D$).

 It should be noted that the usage of the fragmentation function assumes the
absence of the interaction of the produced heavy quark $Q$ with the 
remnants of the initial hadrons. Therefore, it should be no difference between
the spectra of charmed and anticharmed hadrons. Moreover, any modification of
the fragmentation mechanism can not reproduce the production
asymmetry (the leading particle effect).

Note, that the fragmentation mechanism can be apply for the 
production of the $c \bar c$ pair in the color--singlet  state or for high 
$p_T$ production of the open charm. On the other hand, for the case of the 
hadronic production of color $c \bar c$ pair with
small $p_T$ one should takes into account the possibility of charmed
$c$ and $\bar c$ quarks interaction with the initial hadron remnants. 
Therefore, due to the different valence quarks in the initial hadrons 
one may expect the different inclusive spectra of the final charmed hadrons.

In the parton model framework, a heavy $c$--quark should interact with a 
high probability with its nearest neighbor in the rapidity space able to
form a color-singlet state with it. In some cases, the heavy
antiquark may find itself close (in rapidity space) to a valence
light quark from the initial hadron. This would result in the formation
of a fast heavy meson in the fragmentation region of the initial
hadron. Alternatively, the proximity of a heavy quark to
a valence diquark results in the fast charmed $B(c q_1 q_2)$-baryon
production.

Therefore, the "hard" part of charmed hadron spectra is very sensitive to the
form of valence quark distribution in the initial hadrons.

In this note we consider the charmed hadron production in high-energy beam
of $\Sigma^-$-hyperon. We may expect the different behavior of the 
distributions of the valence  $d$- and $s$-quarks. As a result we should
observe the different $x_F$-dependence of spectra of charmed hadrons with
$d$- or  $s$-quarks, namely, $D^-(\bar c d)$ and $D_s^-(\bar c s)$,
$\Xi_c^0(cds)$ and $\Sigma_c^-(cdd)$, etc.

Indeed, very roughly, the distribution of valence quark in the baryon
$B(q_1 q_2 q_3)$ can be presented as follows~\cite{ls1, Likhoded:1997bm}:
\begin{equation}
V^B_{q_1}(x) \propto  x^{-\alpha_1} (1-x)^{\gamma_b - \alpha_2 - \alpha_3},
 \label{vq}
\end{equation}
where $\alpha_i$ is the intercept of the leading Regge-trajectory for
$q_i$-quark, while $\gamma_B \simeq 4$. Note, that due to violation flavor
$SU(N)$-symmetry, we have different intercepts  for
$d(u)$- and $s$-quarks~\cite{collin,klp}:
\begin{eqnarray}
\alpha_u = \alpha_d = \frac{1}{2}, \quad \alpha_s \approx 0, \quad
  \alpha_c \approx -2.2. \label{rec22}
\end{eqnarray}
As a result, the $x$-dependence of the valence $d$ and $s$-quark in the
$\Sigma^-(sdd)$-hyperon has the form as follows~\cite{ls1}:
\begin{equation}
V^{\Sigma}_d \sim \frac{1}{\sqrt{x}} (1-x)^{3.5}, 
\quad
V^{\Sigma}_s \sim (1-x)^{3}    \label{vds}
\end{equation}
It is seen from the~(\ref{vds}) that the valence $s$-quark in the 
$\Sigma^-$-hyperon has slightly harder $x$-distribution than that for 
$d$-quark.

We use the model \cite{ls1, Likhoded:1997bm} to describe the production 
asymmetry for charmed hadrons. 
In this model the interaction of the charmed quarks with valence
quarks from the initial hadrons describes with the help of the recombination 
function \cite{ls1, Likhoded:1997bm}. 
The detail description of this
mechanism is given elsewhere~\cite{ls1, Likhoded:1997bm}. The recombination 
of $q_V$ and 
$\bar c$ quarks into $D$-meson is described by the function of 
$R_M(x_V, z; x)$:
\begin{eqnarray}
 R_M(x_q, z; x) = Z_M 
 \xi_q^{(1-\alpha_q)} \xi_c^{(1-\alpha_c)} \; \delta(1 - \xi_q - \xi_c), 
\label{rec1} 
\end{eqnarray}
where $\xi_q = x_q / x$ and $\xi_c = z / x$, while $x_q$, $z$, and $x$ are the
fractions of the initial-hadron c.m. momentum that are carried away by the
valence $q$-quark, charmed $c$-quark, and the meson $M_{\bar c q}$, 
respectively.
The corresponding recombination of three quarks into baryon can be described
by means of the similar recombination function:
\begin{eqnarray}
 R_B(x_1, x_2, z; x) = Z_M 
 \xi_1^{(1-\alpha_1)}  \xi_2^{(1-\alpha_2)} \xi_c^{(1-\alpha_c)} \; 
\delta(1 - \xi_1 - \xi_2 - \xi_c).
\label{rec2} 
\end{eqnarray}
These functions take into account the momentum  conservation and the 
proximity of partons in the rapidity space. Actually,
the recombination function is the modulus squared of the heavy meson 
wave function in momentum space, being considered in the infinite momentum 
frame in the valence quark approximation. 

As a result, the total differential cross section for the production of 
$H_c$--hadron looks as follows:
\begin{eqnarray}
 \frac{d \sigma(H_c)}{dx} =  \frac{d \sigma^R(H_c)}{dx} +  \; 
   \frac{d \sigma^F(H_c)}{dx}, \label{sig4}
\end{eqnarray}
where the first term in the r.h.s. is the cross section for $H_c$--hadron
production due to the recombination of the charmed $c$--quark with valence 
quarks from initial hadron, while the second term  is the cross section for 
$H_c$ production due to charmed quark fragmentation.

Note, that this model provides more or less successful description of the 
charmed $D$-meson production in $\pi^- \, N$-interactions
(see Fig.~1,~2 
and \cite{Likhoded:1997bm} for details).

We use LO formulas for the cross sections of quark-antiquark and gluon-gluon
annihilation into charmed quark pair. We set $m_c=1.25$~GeV, $\alpha_s=0.3$
and find the following cross section value of $\Sigma^- \, p$ interaction
at $P_{LAB} = 600$~GeV:
\begin{equation}
 \sigma(\Sigma^- \, p \, \to \, c \, \bar c \, X) \simeq 8\,\,\mu{b}
\end{equation}
In our calculations we do not pretend to reproduce the absolute value of
this cross section (see~\cite{1}, for detail consideration of this problem).
We concentrate on the description of $x_F$-distribution of charmed mesons and
baryons. The corresponding distributions (integrated over $p_\top$) are 
presented in Fig.3--5. We may see from these figures, that the considered
charmed quark interaction in the final state (recombination)
leads, indeed, to noticeable differences in $x_F$-spectra. These differences
can be explicitly seen in Fig.6, where we present the corresponding asymmetry
$A$ (see~(\ref{asym} for definition).
The most non-trivial prediction of the proposed model is presented in two
lower plots in Fig.6,
where we present the ratio of the inclusive spectra of $D^{*-}_s(\bar c s)$
and $D^{*-}(\bar c d)$ mesons. It is evident from this figure  the
difference of $x_F$-spectra of the these two mesons, which is a result of of
different $x$-distributions of the valence $d$-and $s$-quarks in the initial
$\Sigma^-$-beam (see~(\ref{vds})). 

\vspace*{0.5truecm}
\noindent {\bf Conclusion} \\
In the present note we wish to stress once more, that the source of the 
observed
asymmetry in charmed hadron production is the interaction of produced charmed
quarks with valence quarks from initial hadrons. Note,
the model under consideration provides also the additional method
to measure the valence
quark distribution functions of $K$-meson and $\Sigma$-baryons.

\newpage
\begin{figure}[t]
\epsfig{file=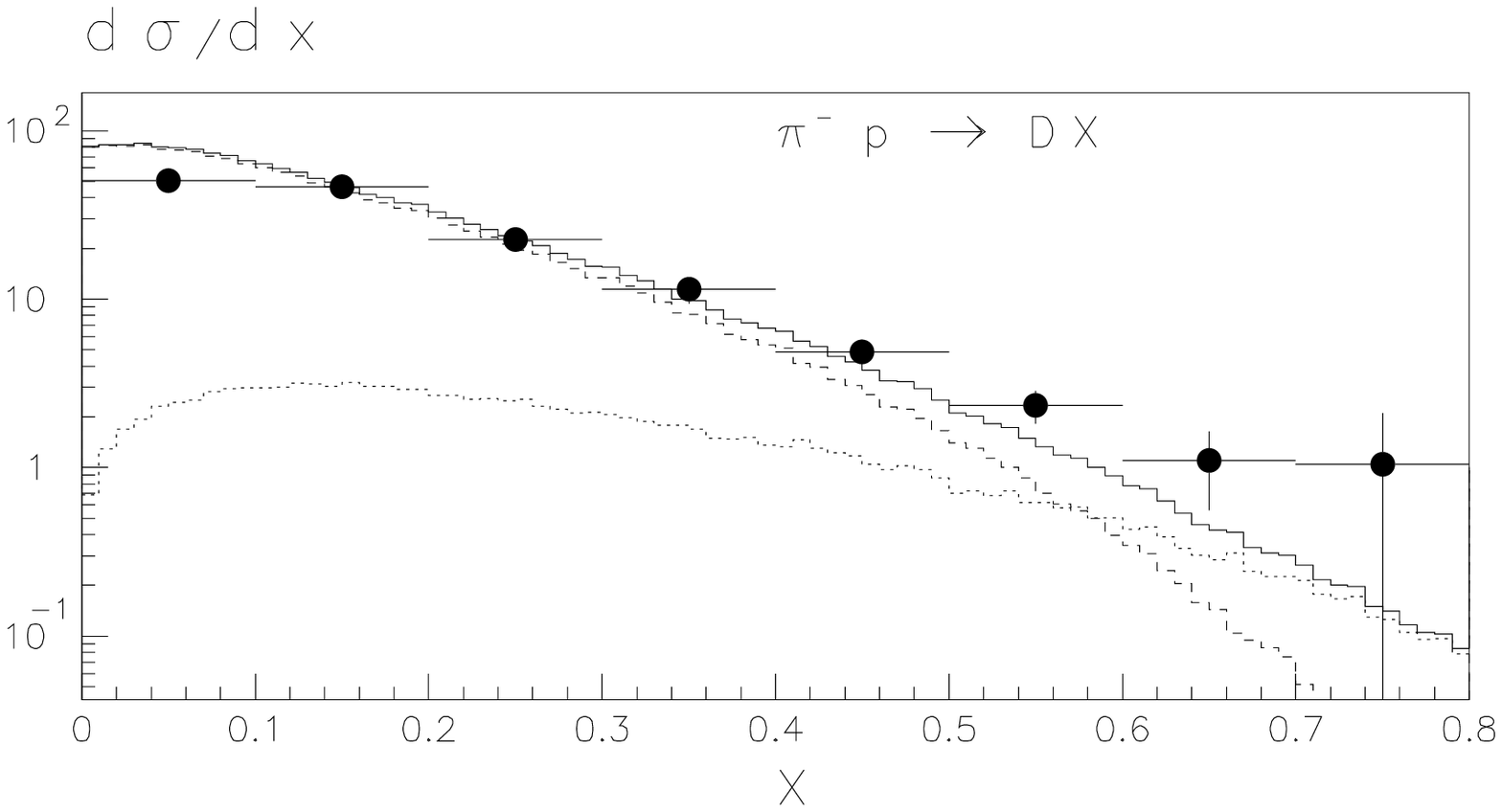,width=14cm,clip} 
\ccaption{}{
Differential distributions $\frac{d \sigma}{dx}$ for
the energy of $E_{\pi} = 250$~GeV. The experimental data are taken 
from~\cite{exp1}. 
The dotted (dashed) histogram corresponds to the recombination (fragmentation)
contribution. The solid histogram represents their sum. The cross sections are
presented in $\mu$b (see~\cite{Likhoded:1997bm}
for details).  }  

\epsfig{file=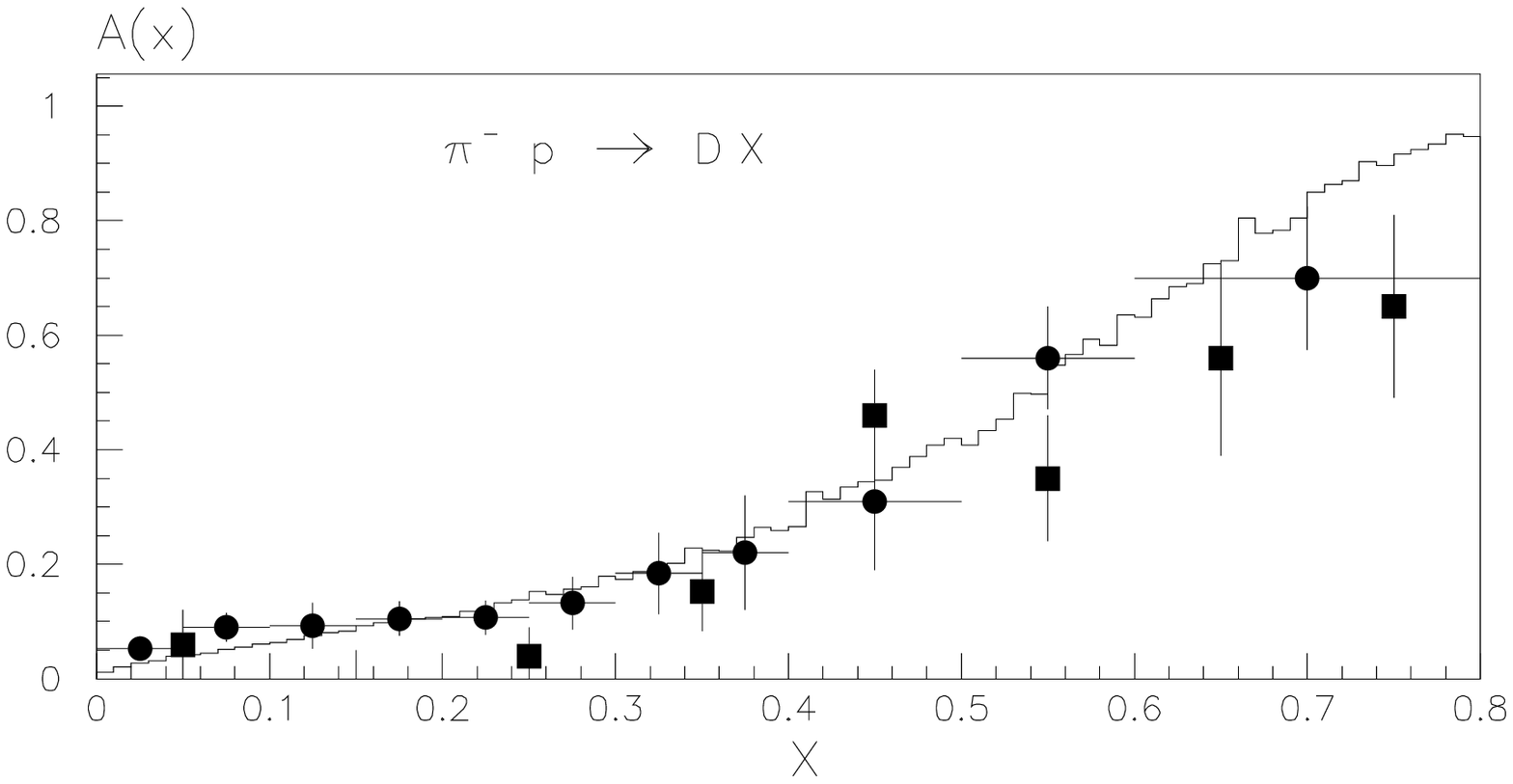,width=14cm,clip} 
\ccaption{}{
The description of the asymmetry $A(x)$ 
in $\pi^- p $ collisions~\cite{exp1,exp2} (see~\cite{Likhoded:1997bm}
for details) } 
\end{figure}

\newpage

\begin{figure}[t]
\epsfig{file=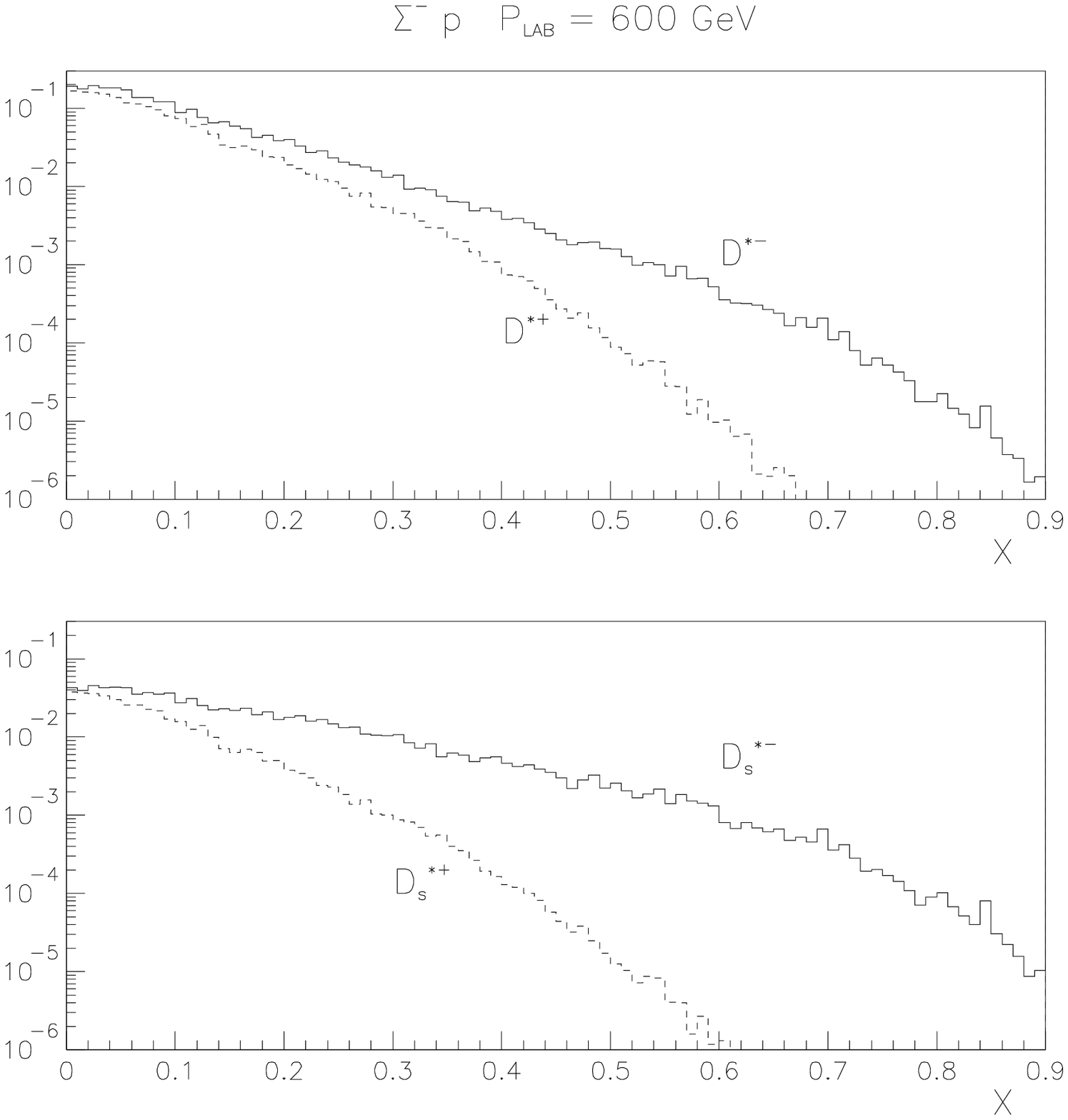,width=16cm} 
\ccaption{}{
Differential distributions of $D^{* \pm}$ and $D_s^{* \pm}$ mesons
produced in $\Sigma^- \, p$ interactions at $P_{LAB} = 600$~GeV.
 } 
\end{figure}

\newpage

\begin{figure}[t]
\epsfig{file=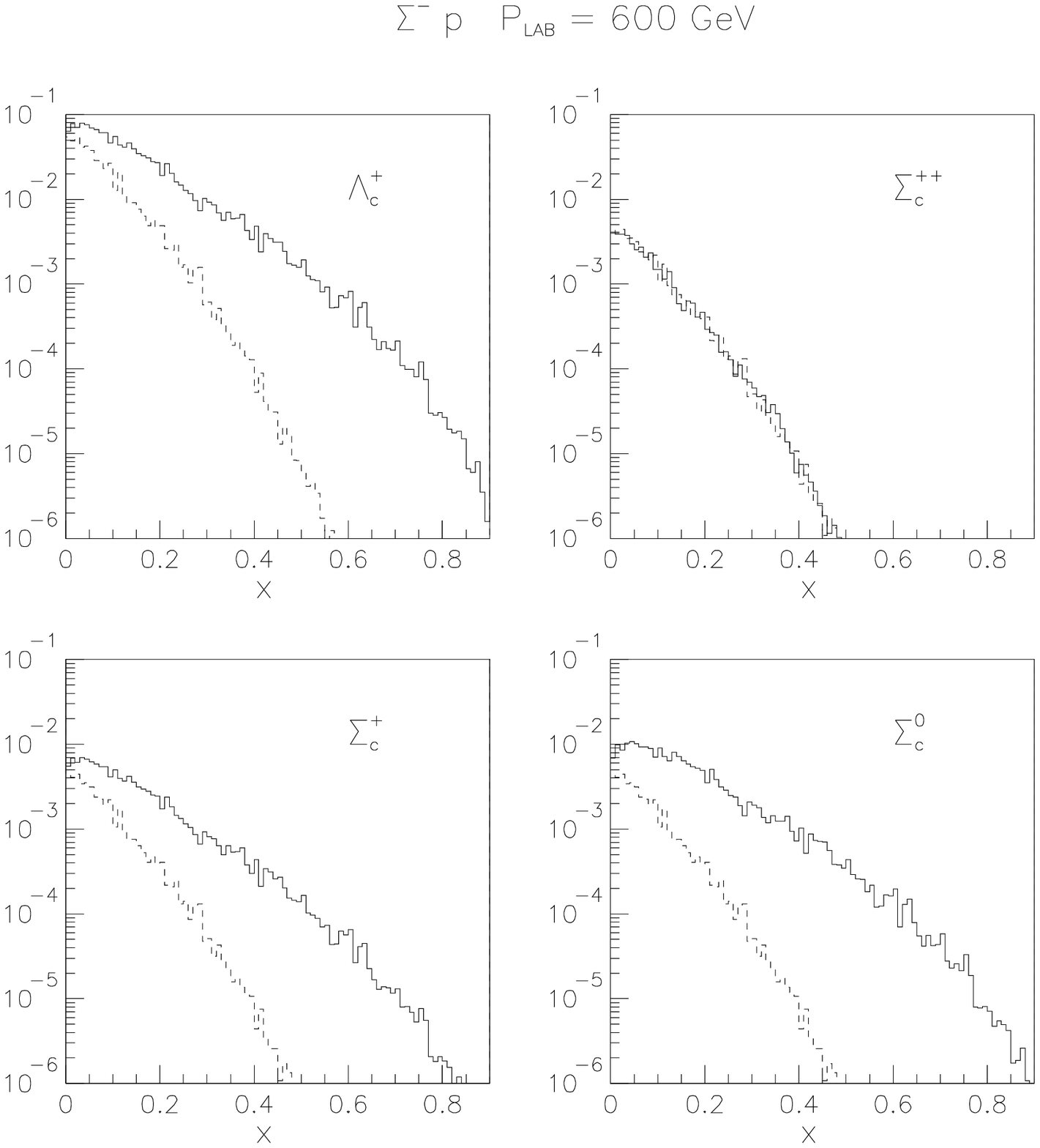,width=16cm} 
\ccaption{}{
Differential distributions of charmed baryons, 
produced in $\Sigma^- \, p$ interactions at $P_{LAB} = 600$~GeV.
The solid~(dashed) curves correspond to   baryon (antibaryon) distributions.
} 
\end{figure}

\newpage

\begin{figure}[t]
\epsfig{file=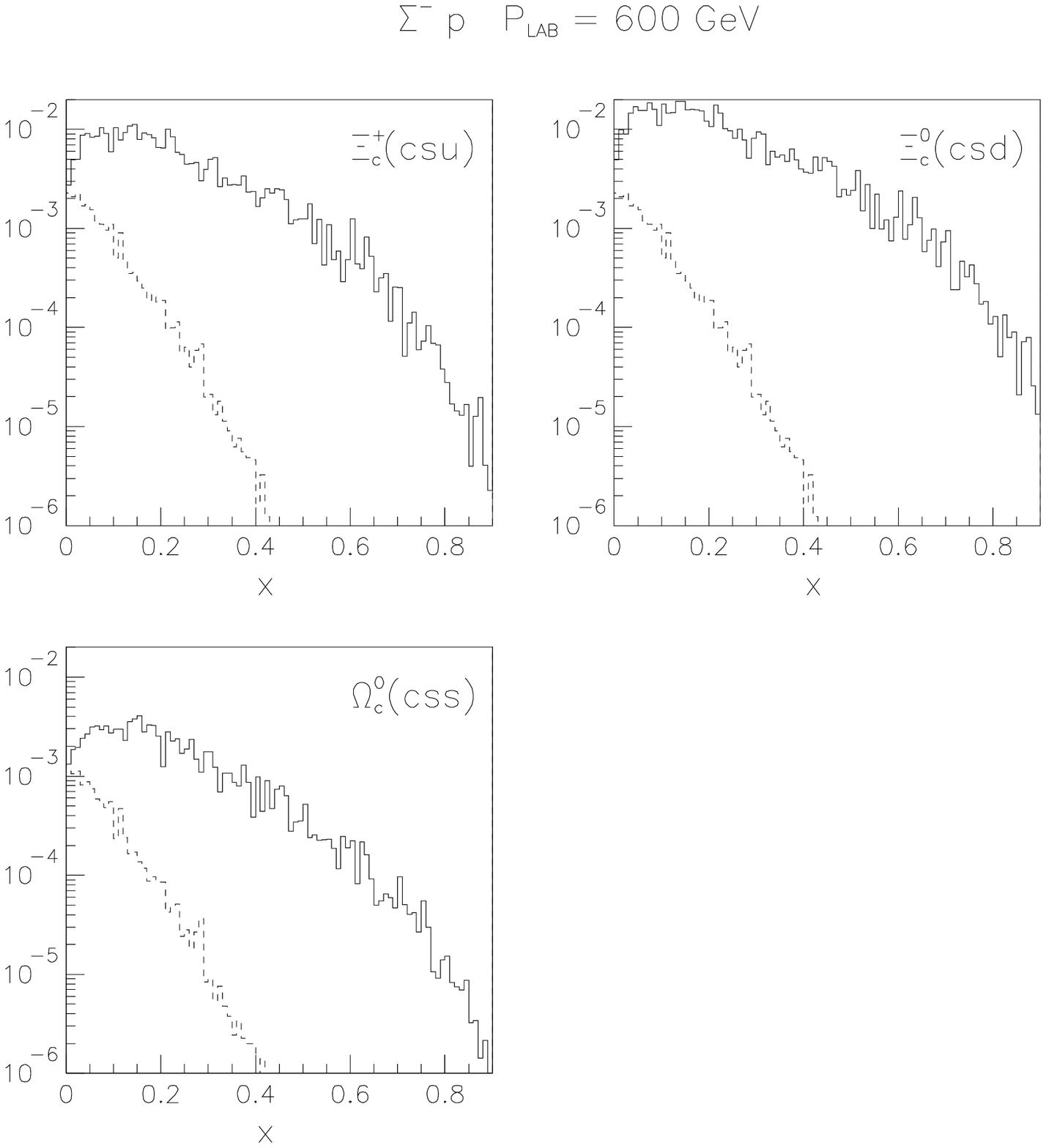,width=16cm} 
\ccaption{}{
The differential distributions of charmed-strange baryons , 
produced in $\Sigma^- \, p$ interactions at $P_{LAB} = 600$~GeV.
The solid~(dashed) curves correspond to   baryon (antibaryon) distributions.
} 
\end{figure}

\newpage

\begin{figure}[c]
\epsfig{file=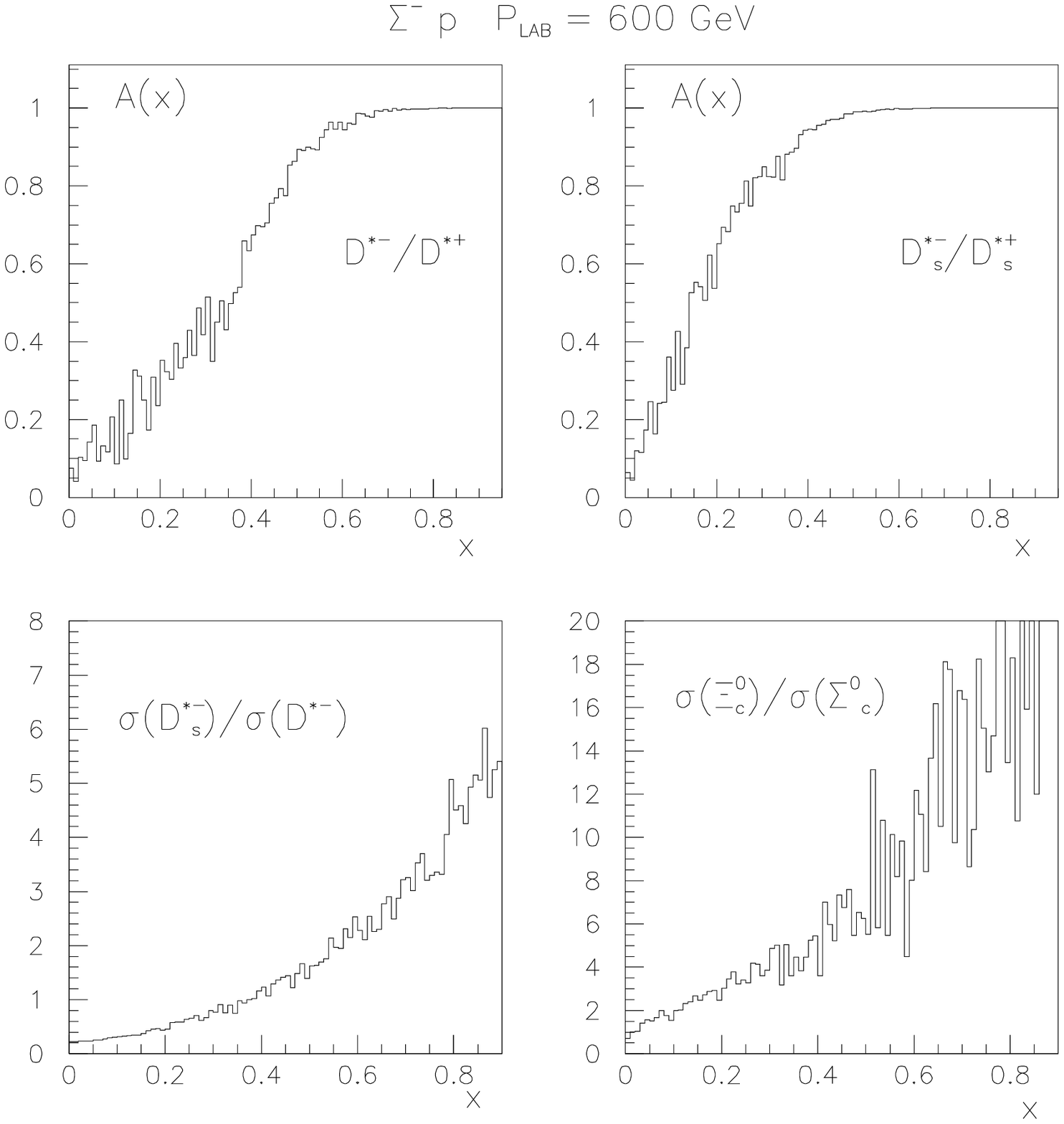,width=16cm} 
\ccaption{}{
Production asymmetry for $D^{* \pm}$ and $D_s^{* \pm}$ mesons (two upper
plots), produced in $\Sigma^- \, p$ interactions at $P_{LAB} = 600$~GeV.
Two lower plots present the ratio of the spectra
of charmed hadrons with and without strange quarks
(i.e. the $D^{* -}_s / D^{* -}$ and 
$\Xi^{0}_c(cds) / \Sigma^0_c(cdd)$ ratios).
} 

\end{figure}

\end{document}